\documentstyle[12pt]{article}

\begin{document}
\begin{titlepage}

\title{The vacuum energy density in the teleparallel
equivalent of general relativity}

\author{J. W. Maluf$\,^{*}$ \\
Instituto de F\'{\i}sica, \\
Universidade de Bras\'{\i}lia\\
C. P. 04385 \\
70.919-970 Bras\'{\i}lia DF, Brazil\\
and\\
J. F. da Rocha-Neto$\,^{\S}$\\
Instituto de F\'{\i}sica Te\'orica,\\
Universidade Estadual Paulista,\\
Rua Pamplona 145,\\
01405-900 S\~ao Paulo SP, Brazil\\}
\date{}
\maketitle

\begin{abstract}
Considering the vacuum as characterized by the presence of only
the gravitational field, we show that the vacuum energy density of
the de Sitter space, in the realm of the teleparallel
equivalent of general relativity, can acquire arbitrarily high
values. This feature is expected to hold in the consideration of
realistic cosmological models, and may possibly provide a simple
explanation the cosmological constant problem.

\end{abstract}
\thispagestyle{empty}
\vfill
\noindent PACS numbers: 98.80.-k, 04.20.-q, 04.20.Cv, 04.20.Fy\par
\bigskip
\noindent (*) e-mail: wadih@fis.unb.br\par
\noindent (\S) e-mail: rocha@ift.unesp.br \par
\end{titlepage}
\newpage
%\doublespace

\noindent

\section{Introduction}
%\noindent {\bf I. Introduction}\par
\bigskip
Recent cosmological data indicate that the Universe is dominated
by a form of exotic energy, the so called dark energy, which is
expected to account for roughly seventy percent of the energy
density of the universe. A strong, widely acceptable possibility
is that this form of energy is due to Einstein's cosmological
constant $\Lambda$ (see, for example, Ref. \cite{Peebles}), which
is presently considered to be of the order
$\Lambda\simeq 10^{-56} cm^{-2}$. This
fact has revived interest in the cosmological constant problem
(CCP)\cite{Zeldovich,Straumann,Ellwanger,Weinberg}, which is
considered by many as one of the most pressing conceptual
problems in physics.

The CCP is characterized by the enormous difference in the expected
values of the vacuum energy density that arises (i) from estimates
in the realm of quantum field theory, namely, by considering the
zero-point energy of quantum fields, and (ii) from estimates of the
vacuum energy density in the context of Einstein's general relativity
with a cosmological constant. If the space-time is endowed with a
cosmological constant, the energy-momentum tensor
acquires the additional term

$$T^\Lambda_{\mu\nu}={\Lambda\over{8\pi G}}g_{\mu\nu}\;.\eqno(1)$$

\noindent It is then asserted that the cosmological constant
is expected to contribute to the vacuum energy density
according to\cite{Zeldovich,Ellwanger,Weinberg}

$$\rho_\Lambda={\Lambda\over{8\pi G}}\simeq  10^{-29}\;
{g\over{cm^3}}\;.\eqno(2)$$

\noindent Alternatively, it is normally considered that the vacuum
energy density acts like a cosmological constant.

Although it is very difficult to calculate the
precise value of the vacuum energy density in quantum field theories,
it is possible to make reasonable estimates of the contributions
of the various fields. One concludes that the quantum vacuum
energy density $\rho_q$ is roughly of the order\cite{Ellwanger}

$$\rho_q\simeq 10^{97}\;{g\over {cm^3}}\;,\eqno(3)$$

\noindent which is 120 orders of magnitude larger than the estimate
given by Eq. (1). Such huge difference constitutes the CCP. An usual
formulation of the CCP is ``Why is the vacuum energy, as given by Eq.
(1), so small?"

In this article we suggest that the CCP may possibly have a simple
explanation in the context of the Hamiltonian formulation of the
teleparallel equivalent of general relativity
(TEGR)\cite{Maluf1,Maluf2}.
We will consider the definition of gravitational energy that
arises in the realm of the TEGR, and that has been applied to several
configurations of the gravitational field, yielding consistent
results in all applications\cite{Maluf2,Maluf3}. In order to
investigate the space-time vacuum energy density in the framework of
the TEGR, we assume that the vacuum is characterized by the
presence of only  the gravitational field. Therefore the energy
density of the vacuum is the energy density of the corresponding
gravitational field.
We will address the de Sitter space-time, which is a rather
simplified model of a universe. Therefore the
analysis to be described below should be understood as a mechanism
that may be applied to the analysis of realistic cosmological
models.\par
\bigskip
\bigskip

\section{The gravitational energy definition of the TEGR}

\bigskip
The TEGR is constructed out of tetrad fields $e_{a\mu}$, where
$a$ and $\mu$ are SO(3,1) and space-time indices, respectively, and
provides a suitable framework where the notion of gravitational
energy can be discussed and easily applied to any space-time
geometry that admits a 3+1 foliation. The constraint equations
of the Hamiltonian formulation of the TEGR are interpreted
as energy, momentum and angular momentum equations for the
gravitational field\cite{Maluf2}.

The Lagrangian density of the TEGR is given by

$$L(e)\;=\;-k\,e\,\biggl( {1\over 4} T^{abc}T_{abc} +
{1\over 2}T^{abc}T_{bac}-T^aT_a\biggr)\;,\eqno(4)$$

\noindent where $e=det(e^a\,_\mu)$,
$T_{abc}=e_b\,^\mu e_c\,^\nu T_{a \mu \nu}$,
$T_{a \mu \nu}=\partial_\mu e_{a\nu}-\partial_\nu e_{a\mu}$,
and the trace of the torsion tensor is given by 
$T_b=T^a\,_{ab}\;.$
The Hamiltonian is obtained by just rewriting the Lagrangian
density in the form $L=p\dot q -H$.
Since there is no time derivative of $e_{a0}$ in (2.1), the
corresponding momentum canonically conjugated $\Pi^{a0}$ vanishes
identically. Dispensing with surface terms the total Hamiltonian
density reads\cite{Maluf1} 

$$H(e_{ai},\Pi^{ai})
=e_{a0}C^a+\alpha_{ik}\Gamma^{ik}+\beta_k\Gamma^k\;,\eqno(5)$$

\noindent where $\lbrace C^a, \Gamma^{ik}$ and $\Gamma^k\rbrace$
constitute a set of primary constraints, and
$\alpha_{ik}$ and $\beta_k$ are Lagrange  multipliers.
Explicit details are given in Ref. \cite{Maluf1}.
The first term of the constraint $C^a$ is given by a total
divergence in the form
$C^a=-\partial_k \Pi^{ak}+\cdot\cdot\cdot\; $.
The equation $C^a=0$ is interpreted as an energy-momentum equation
for the gravitational field of the type ${\sl H^a}-P^a=0$, and
we identify the total divergence on the three-dimensional
spacelike hypersurface as the energy-momentum density of the
gravitational field. The total energy-momentum is defined by

$$P^a=-\int_V d^3x\,\partial_i \Pi^{ai}\;,\eqno(6)$$

\noindent where $V$ is an arbitrary space volume. It is invariant
under coordinate transformations on the spacelike manifold, and
transforms as a vector under the global SO(3,1) group.

The gravitational energy $E_g$ enclosed by an arbitrary space
volume $V$ is defined by\cite{Maluf2}

$$E_g=-\int_V d^3x\,\partial_i \Pi^{(0)i}\;,\eqno(7)$$

\noindent where $\Pi^{(0)i}$ is the momentum canonically conjugated
to $e_{(0)i}$ (Latin indices $i,j,...$ run from 1 to 3). It reads

$$\Pi^{ak}\;=\;k\,e\biggl\{ 
g^{00}(-g^{kj}T^a\,_{0j}-
e^{aj}T^k\,_{0j}+2e^{ak}T^j\,_{0j})$$

$$+g^{0k}(g^{0j}T^a\,_{0j}+e^{aj}T^0\,_{0j})
\,+e^{a0}(g^{0j}T^k\,_{0j}+g^{kj}T^0\,_{0j})
-2(e^{a0}g^{0k}T^j\,_{0j}+e^{ak}g^{0j}T^0\,_{0j})$$

$$-g^{0i}g^{kj}T^a\,_{ij}+e^{ai}(g^{0j}T^k\,_{ij}-
g^{kj}T^0\,_{ij})-2(g^{0i}e^{ak}-g^{ik}e^{a0})
T^j\,_{ji} \biggr\}\;.\eqno(8)$$

\noindent With appropriate boundary conditions expression (7)
yields the ADM energy. This expression satisfies the main
requirements for a gravitational energy definition\cite{Maluf2}.

The torsion tensor cannot be
made to vanish at a point by a coordinate transformation. This
fact refutes the usual argument against the nonlocalizability of the
gravitational energy, and which rests on the reduction of the metric
tensor to the Minkowski metric tensor at a point in space-time by
means of a coordinate transformation.
In any small neighborhood of space the gravitational field
can be considered constant and uniform. The principle of
equivalence asserts that in such neighborhood it is always possible
to choose a reference frame in which the gravitational effects  
are not observed. Thus in such reference frame we should not
detect any form of gravitational energy. Therefore it is reasonable
to expect that the localizability of the gravitational energy depends
on the reference frame, but not on the coordinate system. In fact any
other form of relativistic energy depends on the reference frame.
It turns out that the gravitational energy definition given by
Eq. (7) displays the feature discussed above, namely, it
depends on the reference frame. More precisely,
it depends on the choice of a global set of tetrad fields since the
energy expression is not invariant under local SO(3,1)
transformations of the tetrad field, but is invariant
under coordinate transformations of the three-dimensional spacelike
hypersurface (reference frames are better conceived in terms of
fields of vector bases \cite{Aldrovandi}).

\section{The de Sitter space-time}

The present investigation of the vacuum energy density is based on a
previous analysis of the gravitational energy of the de Sitter
space-time, and rests on two premises. {\bf I.}
If the quantum vacuum has indeed a physical reality, then such an
enormous energy density must produce a gravitational field and a
corresponding enormous gravitational energy density. If, in addition,
the space-time is endowed with a positive cosmological constant, then
such gravitational field is expected to give rise to the repulsive
force typical of the de Sitter space-time. {\bf II.} Since the
vacuum is characterized by the existence of only the gravitational
field, the energy density of the vacuum is the energy density of the
(de Sitter) gravitational field. From this point of view, the energy
density given by Eq. (2) does not represent the actual vacuum energy
density induced by the cosmological constant.

We will show that in the framework of the TEGR it is possible
to arrive at arbitrarily high values of the gravitational energy
density if the space-time is endowed with a positive cosmological
constant. We will consider the de Sitter space-time as just a model
for our discussion. In realistic cosmological models the feature to be
described below should also take place. The
metric tensor for de Sitter space-time is given by

$$ds^2=-\biggl( 1-{{r^2}\over{R^2}}\biggr)dt^2+
\biggl( 1-{{r^2}\over{R^2}}\biggr)^{-1}dr^2+
r^2d\theta^2+r^2\sin^2\theta d\phi^2\;,\eqno(9)$$

\noindent where $R=\sqrt{3\over \Lambda}$ (the discussion presented
below can also be carried out in the context of the
Schwarzschild-de Sitter solution). The de Sitter space-time will be
described by the following set of tetrad fields,

$$e^a\,_\mu=\pmatrix{\alpha^{-1}&0&0&0\cr
0& \alpha\sin\theta\,\cos\phi& r\cos\theta\,\cos\phi&
-r\sin\theta\,\sin\phi\cr
0& \alpha\sin\theta\,\sin\phi & r \cos\theta\,\sin\phi &
r\sin\theta\,\cos\phi\cr
0& \alpha\cos\theta & -r\sin\theta & 0\cr}\;,\eqno(10)$$

\noindent where

$$\alpha=\biggl( 1-{{r^2}\over{R^2}}\biggr)^{-{1\over 2}}\;.$$

The set of tetrad fields given above satisfy the conditions

$$e_{(i)j}=e_{(j)i}\;,\eqno(11)$$

$$e_{(i)}\,^0=0\,.\eqno(12)$$

\noindent In a space-time determined by a global set of tetrad
fields $e^a\,_\mu$ we may consider the existence of an underlying
reference space-time with coordinates $q^a$ that are anholonomically
related to the physical space-time coordinates by means of the
relation $dq^a=e^a\,_\mu dx^\mu$. Conditions (11) and (12) above
on the tetrad fields establish a  unique reference space-time
that is neither related by a boost transformation, nor rotating
with respect to the physical space-time\cite{Maluf2}. Therefore
we will evaluate the gravitational energy of the de Sitter
space-time with respect to the frame defined by Eqs. (11) and (12).

If the tetrad fields satisfy the time gauge condition (12), then
expression (7) reduces to (see Eq. (3.11) of Ref.
\cite{Maluf2})

$$-\int_V d^3x\,\partial_i\Pi^{(0)i}=
{1\over{8\pi G}}\int_V d^3x \partial_j (eT^j)=
{1\over{8\pi G}}\int_S dS_j(eT^j)\;.\eqno(13)$$

\noindent
We have the definitions
$e=\det (e_{(i)j})$, $T^i=g^{ik}T_k=g^{ik}e^{(m)j}T_{(m)jk}$,
and $T_{(m)jk}=\partial_j e_{(m)k}-\partial_k e_{(m)j}$. All these
field quantities are restricted to the three-dimensional spacelike
hypersurface.
As a consequence, the calculation of Eq. (7) out of tetrads given
by Eq. (6) yields precisely the expression obtained in Ref.
\cite{Maluf3}. By integrating Eq. (7) over a surface $S$ defined
by $r=R$ we obtain the total energy\cite{Maluf3}

$$E_{dS}={1\over {8\pi G}}\int _S d\theta d\phi\, (eT^1)
={1\over G}R ={1\over G}\sqrt{3\over \Lambda}\;,\eqno(14)$$

\noindent and the average energy density

$${{E_{dS}}\over{2\pi^2R^3}}={\Lambda \over {6\pi^2 G}}\;.\eqno(15)$$

\noindent In Eq. (14) we have

$$eT^1=2r\,\sin\theta\biggl[1-\sqrt{1-{r^2\over R^2}}\biggr]\;.
\eqno(16)$$

The gravitational energy density is given by
$(1/8\pi G)\partial_r (eT^1)$. By integrating this expression
in $\theta$ and $\phi$ we obtain the gravitational energy per unit
radial distance $\varepsilon(r)$. Therefore upon integration 
in $\theta$ and $\phi$ and differentiation in $r$ of Eq. (16) we
arrive at\cite{Maluf3}

$$E_{dS}=\int_0^R dr \varepsilon(r)\;,\eqno(17)$$

\noindent where 

$$\varepsilon(r)={1\over G}\biggl[ 1+{{2\beta^2 -1}\over
{\sqrt{1-\beta^2}}} \biggr] \;,\eqno(18)$$

\noindent  and $\beta^2= r^2/R^2$.

A crucial point of the present analysis is that $\varepsilon(r)$
is clearly divergent in the limit $r \rightarrow R$, i.e.,
$\varepsilon(r) \rightarrow \infty$. This fact is an indication
that the CCP may have an explanation in the present framework.
In order to verify this issue, let us first note that
$ \varepsilon (r)\,dr$ is the gravitational energy contained within
the spherical shell of radius $r$ and width $dr$. Since the area
of such shell is $4\pi r^2$, the gravitational energy per
unit volume is given by $\rho (r)=\varepsilon(r)/(4\pi r^2)$
(we arrive at precisely the same expression by a suitable
coordinate transformation).
We obtain the correct dimensions upon the introduction of the
velocity of light $c$ by means of the replacement
$1/G \rightarrow c^4/G$. Therefore the gravitational energy
density $\rho(r)$ is given by

$$\rho(r)={c^4\over G} {1\over {4\pi r^2}}
\biggl[ 1+{{2\beta^2 -1}\over
{\sqrt{1-\beta^2}}} \biggr] \;.\eqno(19)$$

\noindent In order to compare Eq. (19)
with expression (3) we use

$${c^4\over G}=1.21 \times 10^{49}\;g{{cm}\over s^2}=
0.135 \times 10^{29} {g\over {cm}}\;.$$

\noindent Therefore we must find a radial position $r$ of an
observer such that

$$(0.135 \times 10^{29} {g\over {cm}}){1\over {4\pi r^2}}
\biggl[ 1+{{2\beta^2 -1}\over
{\sqrt{1-\beta^2}}} \biggr]
\simeq 10^{97}\;{g\over {cm^3}}\;.\eqno(20)$$

\noindent Replacing $r^2$ in the denominator of $1/(4\pi r^2)$  in
Eq. (14) by $\beta^2 R^2=\beta^2 (1.73 \times 10^{28}cm)^2$
we find

$${1\over {\beta^2}} \biggl[ 1+{{2\beta^2 -1}\over
{\sqrt{1-\beta^2}}} \biggr]\simeq 10^{127}\equiv n\;.\eqno(21)$$

\noindent The equation above leads to a simple equation for
$\beta^2$,

$$n^2\beta^4+(4-2n-n^2)\beta^2+2n-3=0\;,\eqno(22)$$

\noindent whose solutions are given by

$$\beta^2={1\over {2n^2}}
\biggl[ -4+2n+n^2 \pm
\biggl(n^4-4n^3+8n^2-16n+16\biggr)^{1\over 2}\biggr]\;.\eqno(23)$$

\noindent The approximate solutions of the equation above are
$\beta^2\simeq 1-1/n^2$, that yields

$$r\simeq R(1-10^{-254})\;,\eqno(24)$$

\noindent and $\beta^2\simeq 2/n$, that implies

$$r\simeq 10^{-63}R\simeq 10^{-35}cm\;.\eqno(25)$$

Therefore in order for an hypothetical observer in the de
Sitter universe to be under the effect of 
the energy density given by Eq. (3), either its position $r$
practically coincides with the cosmological horizon ($r$ given by
Eq. (24)) or the observer must stand
extremely close to the origin of the coordinate system
($r$ given by Eq. (25)).\par
\bigskip

\section{Discussion}

The gravitational energy density given by Eq. (19) may acquire
arbitrarily high values, depending on the value of the coordinate
$r$.
We note that in quantum field theories the establishment of the
ultraviolet cutoff that leads to the value given by Eq. (3) is an
assumption related to the validity of quantum field theory
up to the Planck scale, and is precisely required to avoid a
divergence. In the present context of a classical, simplified
cosmological model we do not yet dispose of a similar argument to
assert that Eq. (19) acquires very large but finite values.
The imposition of cutoffs is a device typical of a quantum field
theories.

The de Sitter space-time describes an empty universe. Inspite of
this fact, if we imagine an observer located at $r\simeq 0$ or
at $r \simeq R$, such observer will experience an enormous
gravitational energy density due to the cosmological constant. In a
{\sl realistic} cosmological model, with an arbitrary matter
distribution, we expect a similar situation to hold, namely, we
expect that there are regions in space where the vacuum energy
density acquires values of the order of Eq. (3). Such energy density
is expected to be related to the quantum vacuum energy
density. According to this picture, the vacuum energy density is
not homogeneous in space. However, the repulsive acceleration in the
de Sitter space-time induced by the cosmological constant,
$a=(1/3)\Lambda r$, also constitutes a
{\sl nonhomogeneous} characteristic of the space-time which,
we believe, is intrinsically related to the above mentioned
nonhomogeneity. 
Even though Eq. (19) holds for a quite simplified model of universe,
the nonhomogeneity of the vacuum energy density (on very large
scales of distance) {\sl cannot} be ruled out
on empirical grounds. On the contrary, this hypothesis may lead to
new ideas about the relation between gravity and quantum theory.

We note finally that the present analysis of gravitational energy
in the de Sitter space is different from the one carried out by
Abbot and Deser\cite{Abbot}, who provided an expression for the
gravitational energy {\sl about} the de Sitter background, i.e.,
they calculated the energy of a field configuration that deviates
from the de Sitter metric and vanish at infinity (this point is
discussed in Ref. \cite{Maluf3}).

In summary, we have presented a simple explanation to
describe the emergence of extremely high values of the vacuum energy
density in the universe. Such mechanism may explain the cosmological
constant problem. The present result is achieved by reinterpreting
the vacuum energy density as the gravitational energy density of the
de Sitter space-time.

\bigskip
\bigskip
\noindent {\sl Acknowledgements}\par
\noindent J. F. R. N. is grateful to the Brazilian agency
FAPESP for financial support.\par
\bigskip
%\bigskip

\end{document}